\documentstyle [12pt] {article}
\begin{document}
\centerline{\Large\bf{Skyrmion mass and a new kind of the cyclotron resonance}}
\centerline{\Large\bf{for 2DEG.}}
\centerline{\quad }
\centerline{S.V.Iordanskii}
\centerline{\small\sl{L.D.Landau Institute for TheoreticalPhysics Rus.Ac.Sci,}}
\centerline{\small\sl{Chernogolovka,Moscow distr.142432, Russia}}
\vskip-13mm
\begin{abstract}
The skyrmionic mass was calculated using gradient expansion method. A special
cyclotron resonance is predicted with the frequency defined by the exchange
energy. The possibility of the extra bound electron is discussed.
\end{abstract}

 The energy and the charge of skyrmion for 2d electron system at high magnetic
field were calculated in the papers [1-4] and [5-6]. The difference in the
energy expression is connected with the interpretation of the kinetic energy
as a constant cyclotron  energy in [2-4] and as a differential operator
in [5-6]. For the experimental investigation of the existence of the
skyrmions it is useful to find specific properties which can be checked by
physical measurements. I discuss here the problem of the skyrmion motion as
a whole what is not directly connected with it's internal energy. For the
calculations I shall use the approach developed in [6,7].

This approach is based upon the transformation of the electron spinors $\psi$
by nonsingular rotation matrix $U(\bf r)$ , $\psi=U\chi$ to new spinors
$\chi$. One get in such a manner Hartree-Fock equation for spinor $\chi$.
 It acquires the form
$$
i\frac{\partial\chi}{\partial t}=\frac{1}{2m}(-i{\nabla}_k-A_{0k}+{\Omega}^l_k
{\sigma}_l)^2\chi-\gamma{\sigma}_z\chi+{\Omega}_t^l{\sigma}_l\chi
$$
  for the simplest case of the local exchange considered in this paper.
Here $\sigma_l$ are Pauli matrices and $-iU^{+}{\partial}_kU=\Omega^l_k
{\sigma}_l$ with
\begin{displaymath}
\vec\Omega^z=\frac{1}{2}(\nabla\alpha+cos\beta\nabla\alpha)
\end{displaymath}
\begin{displaymath}
\vec\Omega^x=\frac{1}{2}(sin\beta cos\alpha \nabla\alpha-sin\alpha\nabla\beta)
\end{displaymath}
\begin{equation}
\label{*}
\vec\Omega^y=\frac{1}{2}(sin\beta sin\alpha \nabla\alpha+cos\alpha\nabla\beta)
\end{equation}
and $\gamma$ is the exchange constant. I assume that Euler angles are
$\alpha,\beta,\alpha$ with two equal angles
 to avoid the singularity
in $\Omega^l$
for nontrivial degree of mappings $Q$ and suppose $cos\beta=-1$ at
the singularity point of $ \alpha(\vec r)$ ([6,7]).
This equation for the electrons in the field of rotation matrix $U$ is fully
equivalent to the nonrotated Hartree-Fock equation with $\Omega^l=0$ but
 the nonuniform exchange term
$-\gamma{\bf n}({\bf r})\mbox{$\boldmath \sigma $}$.
Here ${\bf n}({\bf r})$ is the unit vector in the direction of the mean spin.

I shall assume that the rotation matrix which adjust the spin direction to
a given mean spin direction at any point of 2d plain depends on the position
of the skyrmion center $U=U(\vec r-\vec X)$ and calculate the proper term in
the skyrmion action due to the time dependence of $\vec X(t)$. For the
calculation I must find the electron action for the ground state in the field
of the matrix $U$. In the proper electron hamiltonian I shall have the
additional perturbation term [6,7] $$H_1=iU^{+}\nabla U \vec X_t=-\vec \Omega^l
\sigma_l\vec X_t$$.
                          In order to find the proper term in the skyrmionic
action I must perform perturbation theory calculations in $H_1$.

Due to isotropy of the system there is no linear term in $\vec X_t$ and one
must find the second order term in the action $S=iTr ln G$ where $G$ is the
electronic Green function. The second order term in the action is
\begin{equation}
\label{1}
\delta S=\frac{i}{2}TrH_1G_0H_1G_0
\end{equation}
where $G_0$ is the unperturbed Green function for Hartree-Fock equation
with $\Omega^l=0$
\begin{equation}
G_0(\vec r,\vec r',t-t')=\sum_{s,p}\int \frac{d\omega}{2\pi}e^{i\omega(t'-t)}
g_s(\omega)\Phi_{s,p}(\vec r)\Phi^{+}_{s,p}(\vec r')
\end{equation}.
Here  $\Phi_{s,p}$ are normalized Landau wave functions in Landau gauge and
the summation
is over all $s$ and $p$. Matrices $g_s(\omega)$ correspond to the full
filling of the lowest spin sublevel for $s=0$
\begin{equation}
g_0(\omega)=\frac{1+\sigma_z}{2}\frac{1}{\omega-\omega_c/2+\gamma-i\delta}+
\frac{1-\sigma_z}{2}\frac{1}{\omega-\omega_c/2-\gamma+i\delta}
\end{equation}
and all other states are empty
\begin{equation}
g_s(\omega)=\frac{1}{\omega-\omega_c(s+1/2)+\gamma\sigma_z+i\delta}
\end{equation}
where $\delta\rightarrow(+0)$.

The main term with no derivatives of $\Omega^l$ and $\vec X_t$ in (\ref{1})
corresponds only to $s=0$. Also only cross terms are essential with the
poles in $\omega$ above and under the real axe
\begin{equation}
\delta S=\frac{1}{2}\int Tr(\vec\Omega^l\vec X_t\sigma_l)g_0(\omega)
(\vec\Omega^{l'}\vec X_t\sigma_{l'})g_0(\omega)e^{i\delta\omega}\frac{d\omega}{2\pi}
\frac{d^2r}{2\pi}dt
\end{equation}
I perform here the integration over intermediate space coordinates and
the summation over $p$. It is easy to see that only the terms with $l=l'\ne z$
give non zero contribution. Using the isotropy and performing simple
integration over $\omega$ one gets
\begin{equation}
\delta S=\frac{1}{2\gamma}\sum_{l\ne z}\int \frac{(\Omega^l_x)^2+(\Omega^l_y)^2}
{2}{\dot X}^2\frac{d^2r}{2\pi}dt=\int\frac{{\dot X}^2}{16\gamma}(\frac{\partial n_i}{\partial r_k})^2
\frac{d^2r}{2\pi}dt
\end{equation}
Here I use the expressions (1) for $\Omega^l$ and introduce
the unit vector $$\vec n=(cos\beta,sin\beta cos\alpha,sin\beta sin\alpha)$$.

It is known that for the state with minimal skyrmion energy for the given
degree of mapping $Q$ the value of the space integral [8] is
$$
\frac{1}{2}\int (\frac{\partial n_i}{\partial r_k})^2d^2r=4\pi|Q|
$$
 Therefore the kinetic energy term in the Lagrangian is $E_{kin}=\frac{m{\dot X}^2}{2}$
where $m_s=\frac{|Q|}{2\gamma}$ or in usual units
$$m_s=\frac{eB|Q|}{2c\gamma}$$
where $B$ is the external magnetic field. As it was obtained in a number
of papers (see e.g [3-5]) the skyrmion has the charge $eQ$.
 For the charged skyrmion there are
also linear in $\vec X_t$ terms in the Lagrangian corresponding to the
product of the skyrmion current and the vector-potential of the external
magnetic field $\vec B$
$$
Q\vec X_t\vec A_0
$$
This term can be also calculated by the differentiation of the proper
additional phase of the wave function obtained by the translation of the
skyrmion charge $eQ$. Therefore the full Lagrangian for the motion of the
skyrmion as a whole is (in usual units)
$$
L=\frac{m_s{\dot X}^2}{2}+\frac{e}{c}Q\vec {\dot X}\vec A_0
$$
 The hamiltonian conjugate to $\vec X$ momentum is
 $P_i=\frac{\partial L}{\partial{\dot X}_i}$ which can be considered as a quantum
 operator with usual commutation relations $\left[P_iX_i\right]=i\hbar$.
 Therefore one have the cyclotron energy for the motion of the skyrmion as
 a whole
 $$
 \hbar\omega_s=\frac{eB}{m_sc}=2\gamma
 $$.
 The minimal energy of such motion is
 $$
 \frac{\hbar\omega_s}{2}=\gamma
 $$
 and must be added to the internal energy of skyrmion. In experiments with
 enough number of charged skyrmions one must observe the cyclotron resonance
with
 the frequency defined by exchange energy per electron
 $$
 \omega_s=\frac{1}{\hbar}2\gamma=\frac{e^2}{\hbar l_B}\sqrt{2\pi}
 $$
 where $l_B=\sqrt{\frac{c\hbar}{eB}}$. The final expression is obtained from
 the expression for the exchange energy for fully filled Landau level.

The preceding considerations have some important consequence. The
thermodynamical enrgy for the system with given chemical potential is
the quantum average $<H-\mu N>$ where $H$ is the Hamiltonian, $\mu=(\hbar
\omega_c)/2$ is the chemical potential and $N$ is the particle number.
The change of the total
thermodynamic energy due to the formation of the charged skyrmion is
$E_{tot}=E_{int}+\gamma$ where
$E_{int}$ is the internal energy not including it's motion as a whole in
the external magnetic field. If one put an extra electron in the skyrmion
core it's energy will consist from two main parts. One part is the increase
in the exchange energy $\gamma$ because the added electron must have the
reversed spin direction according to Pauli principle (all lower states are
filled). The other part is the negative Coulomb energy due to the electron
interaction with skyrmion charge $\sim(-e^2Q/L_c)$ where $L_c$ is the
skyrmion core
size. All other terms in the electron energy are comparatively small
$(\sim{1/L_c^2)}$ and can be neglected for the large size of the
core. The added electron make the total skyrmion-electron complex neutral.
Therefore there is no correction to it's energy connected with the motion
of the complex as a whole  in the external magnetic field. The lowest
energy of the complex is $E_{compl}=E_{int}+\gamma-const.e^2Q/L_c$ which is
lower  than the energy of the charged skyrmion for positive $Q$ and $\mu=
(\hbar\omega_c)/2$.
One sees that the skyrmion
with the large size of the core and positive charge must have the bound
electron and become
neutral.The spin direction of this electron is reversed to the direction of the
 average spin in the middle of the core i.e. coincides with the direction of
the mean spin at large distances from the core. The results of [5-6] gives
the negative thermodynamic energy $E_{int}$ because of the strong reduction
of the kinetic energy by $-(\hbar\omega_c)/2$ for $\mu=(\hbar\omega_c)/2$.
Therefore such neutral skyrmions must be spontaneously created.

  The performed calculations use the assumption of the large size of
  skyrmion core otherwise the perturbation theory in $\Omega_l$ is invalid.
The large size of the core requires the small enough g-factor
   (see e.g.[6]). The conclusions must be numerically checked for the real
values of g-factor and magnetically field.

  The reaserch described in this publication was made in part due to award
  RP1-273  of US CRDF for the countries of the FSU. It is also supported
  by Intas grant 95-1/Ru-675.
  

\begin{thebibliography}{99}
  \bibitem[1]{1} S.Sondhi, A.Kahlrede, S.Kivelson, E.Rezayi, Phys.Rev.,B47,
  16418 (1993)
  \bibitem[2]{2} H.Fertig, L.Brey, R.Cote, A.MacDonald, Phys.Rev.,B50,11018,
  (1994)
  \bibitem[3]{3} K.Moon, H.Mori, Kun Yang, S.Girvin, A.Macdonald, Phys.Rev.
  B51,5138(1995)
  \bibitem[4]{4} Yu.Bychkov, T.Maniv, I.Vagner, Phys.Rev.B53, 10148(1995)
  \bibitem[5]{5} S.Iordanskii, S.Plyasunov, Pisma ZhETF,65,253, (1997)
  \bibitem[6]{6} S.Iordanskii, S.Plyasunov, ZhETF v 112(in press),
  cond-mat/9706236 (1997)
  \bibitem[7]{7} S.Iordanskii, Pisma ZhETF v {\bf 66} 3, 178 (1957)
  \bibitem[8]{8} A.Belavin, A.Polyakov, Pisma ZhETF, 22, 245 (1975)
  \end{thebibliography}
\end{document}